\documentclass[10pt,english,a4paper,showpacs,amsmath,amssymb,floatfix,twocolumn]{revtex4-1}
\usepackage{amsmath,amstext,amssymb,graphicx,babel,geometry,setspace,bm,dcolumn,color}
\begin{document}

\title{Early signatures of regime shifts in gene expression dynamics}
\author{Mainak Pal}
\email{mainakpl@bosemain.boseinst.ac.in}
\author{Amit Kumar Pal}
\email{ak.pal@bosemain.boseinst.ac.in}
\author{Sayantari Ghosh}
\email{sayantari@bosemain.boseinst.ac.in}
\author{Indrani Bose}
\email{indrani@bosemain.boseinst.ac.in}
\affiliation{Department of Physics, Bose Institute, 93/1, Acharya Prafulla Chandra Road, Kolkata - 700009, India}

\begin{abstract}
Recently, a large number of studies have been carried out on the early
signatures of sudden regime shifts in systems as diverse as ecosystems,
financial markets, population biology and complex diseases. Signatures
of regime shifts in gene expression dynamics are less systematically
investigated. In this paper, we consider sudden regime shifts in the
gene expression dynamics described by a fold-bifurcation model involving
bistability and hysteresis. We consider two alternative models, Models
1 and 2, of competence development in the bacterial population 
\textit{B.subtilis} and determine some early signatures of the regime shifts
between competence and vegetative state. We use both deterministic and
stochastic formalisms for the purpose of our study. The early signatures
studied include the critical slowing down as a transition point is
approached, rising variance and the lag-1 autocorrelation function,
skewness and a ratio of two mean first passage times. Some of the
signatures could provide the experimental basis for distinguishing
between bistability and excitability as the correct mechanism for
the development of competence.
\end{abstract}


\maketitle

\section{Introduction}
\label{intro}

Complex dynamical systems, ranging from ecosystems and the climate
to gene regulatory networks and financial markets, are known to exhibit
abrupt shifts from one dynamical regime to a contrasting one at critical
parameter values \cite{scheffer091,scheffer092,strange07,scheffer03}. 
Examples of sudden regime shifts include
the collapse of vegetation when semi-arid conditions prevail, the
transition from a clear lake to a turbid one, sudden changes in fish/wildlife
populations \cite{scheffer03,guttal08}, distinct changes in the climate and patterns
of oceanic circulation \cite{scheffer091,lenton11}, financial markets undergoing global
crashes \cite{scheffer091}, spontaneous systemic failures such as asthma attacks
\cite{venegas05}, or epileptic seizures \cite{mcsharry} etc. In a gene regulatory
network, a sudden transition may occur from one stable gene expression
state to another at a critical parameter value \cite{ozbudak,pomerening08,ferrell01,veening08}. 
The induction of the \textit{lac} operon in \textit{E. coli }results in the synthesis
of the protein $\beta$-\textit{galactosidase} required for breaking
up sugar molecules and releasing energy to the cell. There is now
experimental evidence \cite{ozbudak} that an abrupt transition from the uninduced
($\beta$-\textit{galactosidase} level low) state to the induced ($\beta$-\textit{galactosidase}
level high) state of the \textit{lac} operon occurs at a critical value
of an inducer concentration.

\begin{figure}
\includegraphics[scale=0.3]{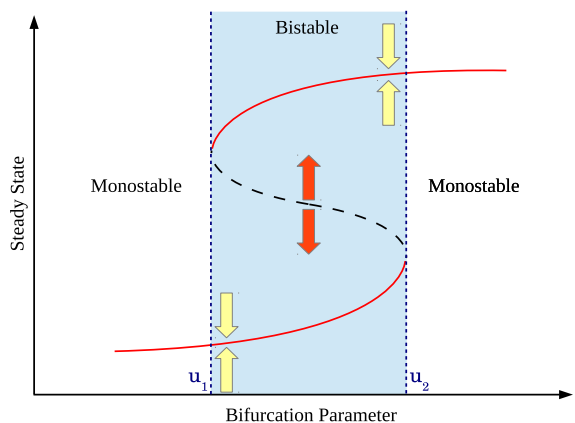}
\caption{A generic steady state versus bifurcation parameter diagram.
The shaded region represents the region of bistability separating
two regions of monostability. The solid lines correspond to stable
steady states and the dashed line represents unstable steady states. The points
$u_{1}$ and $u_{2}$ are the lower and upper bifurcation points respectively.
The arrows indicate the time evolution of a weakly perturbed system
with the system moving towards stable steady states and moving away
from unstable steady states} 
\label{fig1}
\end{figure}

The regime/state shifts in the examples mentioned above are mostly
a consequence of the fold-bifurcation (or the fold-catastrophe), well-characterized
in dynamical systems theory \cite{scheffer091,scheffer092,lenton11,stogatz94}. 
Figure \ref{fig1} illustrates
a specific type of the fold-bifurcation based on bistability and hysteresis,
which provides a physical understanding of the features associated
with sudden regime shifts. The plot represents the steady states of
a dynamical system versus a specific parameter. The state of the dynamical
system at time $t$ is defined by the magnitudes of the relevant variables
at $t$. In the steady state, the net rate of change in the magnitude
of a variable is zero so that there is no further time evolution.
The solid lines in Figure \ref{fig1} denote stable steady states separated
by a branch (dotted line) of unstable steady states. The steady state
curve is folded backwards giving rise to bistability, i.e., the existence
of two stable steady states in the shaded region. The parameter
values $u_{1}$ and $u_{2}$ represent the bifurcation points at which
the abrupt regime shifts from one stable steady state to another occur.
On crossing the bifurcation point, the system loses bistability and
becomes monostable. The transition from one branch of stable steady
states to the other is not reversible but exhibits hysteresis. This
implies that if a transition takes place at a bifurcation point, say
the upper one, the reverse transition from the upper to the lower
branch occurs at the lower bifurcation point and not at the upper
bifurcation point itself. Bistability owes its origin to the presence
of one or more positive feedback loops governing the system dynamics
with the added condition that the dynamics be sufficiently nonlinear 
\cite{pomerening08,ferrell01,veening08}. 
Each stable steady state has its own basin of attraction
in the state space, i.e., the space of all states. In the case of
more than one stable steady state, the location of the initial state
of the system in a basin of attraction determines the steady state
attained by the system in the course of time.

The stability of a steady state indicates that the system returns
to the state on being weakly perturbed from it. This is shown by arrow
directions in Figure \ref{fig1}. If the perturbation is sufficiently strong
so that a transition takes place from one basin of attraction to the
other, a switch occurs between the stable steady states even before
the bifurcation point is reached. Closer the system is to the bifurcation
point, the lesser is the magnitude of the perturbation needed to bring
about the switch. In the example of Figure \ref{fig1}, the branch of unstable
steady states constitutes the border between the two basins of attraction.
The distance of a stable steady state from the corresponding unstable
steady state is a measure of the resilience (robustness against perturbation)
of the system. The resilience is gradually weakened as the system
approaches a bifurcation point. The dynamics of natural systems, in
general, have a stochastic component due to the presence of random
external influences and the inherently probabilistic nature of the
processes involved in the dynamics. Consider the dynamics of gene
expression in a gene regulatory network. Gene expression consists
of two major steps: transcription and translation during which messenger
RNA (mRNA) and protein molecules respectively are synthesized. The
biochemical events (e.g., the binding of a RNA polymerase at the promoter
region of the DNA to initiate transcription) constituting gene expression
are inherently stochastic in character resulting in fluctuations (noise)
around mean mRNA and protein levels \cite{elston05,raj08}. Extrinsic influences
like variability in the number of regulatory molecules also contribute
to the noise. Instead of a single steady state protein level, as in
the case of deterministic time evolution, one now has a steady state
probability distribution in the protein levels reflecting the stochastic
nature of the time evolution. In the presence of low/moderate amounts
of noise, the physical picture underlying sudden regime shifts still
remains valid. As in the case of applied perturbations, fluctuations
of sufficiently strong magnitude can bring about regime shifts before
the bifurcation point is crossed. A number of early signatures of
regime shifts have been proposed so far \cite{scheffer091,lenton11,vannes07,dakos12} in the
scenario of the fold-bifurcation. These are the critical slowing down
(CSD) and its associated effects, namely, rising variances and
the lag-1 autocorrelation function as the critical transition point
is approached, increased skewness in the steady state probability
distribution and the presence of flickering transitions between the
stable steady states. The utility of such signals in the cases of
ecological and financial systems \cite{scheffer091} and complex deteriorating
diseases \cite{chen12} is significant, specially, in developing appropriate
risk-aversion/management strategies. Also, in the absence of physical
models capturing the essential dynamics, one can obtain quantitative
measures of the early signatures by analyzing time-series data.

\begin{figure}
\includegraphics[scale=0.4]{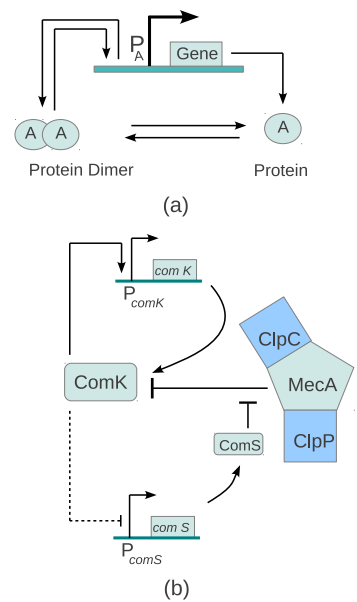}
\caption{Two models, Model 1 (a) and Model 2 (b) of competence development.
(a) The \textit{comK} gene expresses ComK proteins (represented by A) which
form dimers. A dimer binds the promoter region of the gene and activates
the initiation of transcription. The dimer can also unbind from the
promoter region and dissociate into free monomers. (b) The autoregulatory
positive feedback loop mediated by ComK proteins is present. In addition
there is a negative feedback loop in which ComK inhibits the expression
of the \textit{comS} gene and the ComS proteins repress the activity of the
MecA-ClpP-ClpC Complex which targets ComK proteins for degradation.} 
\label{fig2}
\end{figure}
In this paper, we compute quantities providing early signatures of
abrupt regime shifts in gene expression dynamics. We specifically
focus on the phenomenon of competence development in the soil bacteria
\emph{B.subtilis} involving binary gene expression , i.e , the existence
of two expression states. We consider two alternative models of competence
development based on bistable and excitable dynamics. In Section 2, the two models 
are described and their different dynamics highlighted.
In Section 3, we define the quantitative measures of the early signatures
of sudden regime shifts. Section 4 contains the results of our computations
on these early signatures. Section 5 contains concluding remarks. 
 
\section{Models of Competence development}
\label{models}

Microorganisms like bacteria have to cope with a number of stresses during their
lifetime. The bacteria adopt a number of strategies to enhance their
chances of survival under changed circumstances \cite{veening08,collins11}. One
of these is the generation of phenotypic heterogeneity (no genetic
change) in the bacterial population so that a fraction of the population,
even if small, may survive under stressful conditions. In the \textit{B.subtilis} population,
 only a part of the population, in which
the level of a key transcription factor ComK is high, develops competence
\cite{veening08,samoilov06}. The rest of the population is in the so-called vegetative state. 
The ComK protein acts as a master regulator activating the
transcription of several genes including those essential for the uptake
of foreign DNA from the environment. The incorporation of the new
DNA into the bacterial genome confers favorable traits on the bacterial
subpopulation with high ComK level, enabling the subpopulation to
adapt to stress. The ComK activity resulting in the development of
competence is in turn controlled by a host of other proteins. The
core module of the complex regulatory network consists of an autoregulatory
positive feedback loop in which the ComK proteins promote their own
production. The positive feedback gives rise to binary gene expression
in the cell population, i.e., two distinct subpopulations with low
and high ComK levels respectively. Two independent experiments \cite{smits05,maamar05}
provide confirmation that an autoregulatory positive feedback
loop of ComK production is by itself sufficient to establish binary
gene expression in a bacterial population. Considering deterministic
time evolution, the steady state ComK level versus an appropriate
gene expression parameter exhibits a hysteresis curve, resulting from
the fold-bifurcation, similar to the one shown in Figure \ref{fig1} \cite{karmakar07}.
In this scenario, if the cells in a population are prepared to be
in the same initial state, all the cells evolve to the same final
state giving rise to a homogeneous cell population. The generation
of heterogeneity in the form of two distinct subpopulations is brought
about by fluctuation-driven transitions between the low and high ComK
expression states, the fluctuations being associated with the ComK
levels. This gives rise to the experimentally observed \cite{veening08,samoilov06}
bimodal distribution in the ComK levels, i.e., a distribution with
two prominent peaks. There is now experimental evidence \cite{maamar07} that
a lower fraction of the \textit{B.subtilis} population develops competence
with reduced noise in the low ComK level. An alternative physical
mechanism underlying competence development has been proposed by S\"{u}el
et. al. \cite{suel06} in terms of excitability in the dynamics of the genetic
circuit. The excitable core module consists of both positive and negative
feedback loops which bring about a transient activation to the high
ComK state. In this case, there is only one stable steady state (low
ComK level) and two unstable steady states the lower of which, in
terms of the ComK level, sets a threshold for switching \cite{suel06}. Fluctuations
in the low ComK level activate the switch to expression states in
the neighborhood of the state with high ComK level which constitutes
an unstable steady state. The transient activation is followed by
an ultimate return to the stable low ComK state. At any instant of
time, the population divides into two subpopulations with low and
high ComK levels respectively, signifying a different origin for the
bimodal distribution. Quantitative time lapse fluorescence microscopy
provides experimental evidence \cite{suel06} of the probabilistic and transient
differentiation into competence.

We first consider Model 1 in which the protein product
of a single gene autoactivates its own production via a positive feedback
loop. The genetic circuit is shown in Figure \ref{fig2}(a) and, as mentioned
earlier, constitutes a core module of the complex network resulting
in competence development. The proteins synthesized by the \textit{comK
}gene form dimers. The dimer molecules bind at the promoter
region of the DNA and activate gene expression. The gene also synthesizes
proteins at a basal level. A kinetic scheme of the model is as follows
\cite{karmakar07}:
\begin{eqnarray}
 P_{2}+G&\underset{k_{2}}{\overset{k_{1}}{\rightleftharpoons}}&GP_{2}
 \underset{k_{d}}{\overset{k_{a}}{\rightleftharpoons}}G^{*}\nonumber\\
 G&\overset{J_{0}}{\rightarrow}&P\nonumber\\
 G^{*}&\overset{J_{1}}{\rightarrow}&P\nonumber\\
 P+P&\underset{K_{e}}{\rightleftharpoons}&P_{2}\nonumber\\
 P&\overset{k_{p}}{\rightarrow}&\phi
 \label{scheme1}
\end{eqnarray}
\noindent The gene can be in two possible states: inactive ($G$) and active ($G^{*}$).
Proteins are synthesized with rate constant $J_{1}$($J_{0}$)
in the state $G^{*}$($G$) with $J_{0}\ll J_{1}$. The synthesized
proteins dimerize with $K_{e}$ as the equilibrium dissociation constant.
The protein dimer $P_{2}$ binds to the gene in state $G$ and forms
the complex $GP_{2}$ with $k_{1}$ and $k_{2}$ being the rate constants
for binding and unbinding. The complex $GP_{2}$ in turn is activated
to the state $G^{*}$, the rate constants $k_{a}$ and $k_{d}$
being the activation and deactivation rate constants. The synthesized
proteins are degraded with a rate constant, $k_{p}$, $\phi$ denoting
the degradation product. As shown by Karmakar and Bose \cite{karmakar07}, the
kinetic scheme displayed in Equation (\ref{scheme1}) can be mapped onto a simpler scheme
\begin{eqnarray}
G&\underset{k^{\prime}_{d}}{\overset{k^{\prime}_{a}}{\rightleftharpoons}}&G^{*}\nonumber\\
G&\overset{J_{0}}{\rightarrow}&P\nonumber\\
G^{*}&\overset{J_{1}}{\rightarrow}&P\nonumber\\
P+P&\underset{K_{e}}{\rightleftharpoons}&P_{2}\nonumber \\
P&\overset{k_{p}}{\rightarrow}&\phi
 \label{scheme2}
\end{eqnarray}
\noindent The effective activation and deactivation rate constants $k^{\prime}_{a}(x)$
and $k^{\prime}_{d}$ are:
\begin{equation}
k_{a}^{\prime}(x)=k_{a}\frac{(x/k_{s})^{2}}{1+(x/k_{s})^{2}},\;k_{d}^{\prime}=k_{d}
\label{paramrk}
\end{equation} 
\noindent In Equation (\ref{paramrk}), $x$ denotes the protein concentration 
and $k_{s}=\sqrt{\frac{k_{2}}{k_{1}}K_{e}}$.
In the simplified scheme, the rate of change of the protein concentrations:

\begin{figure}
\includegraphics[scale=0.4]{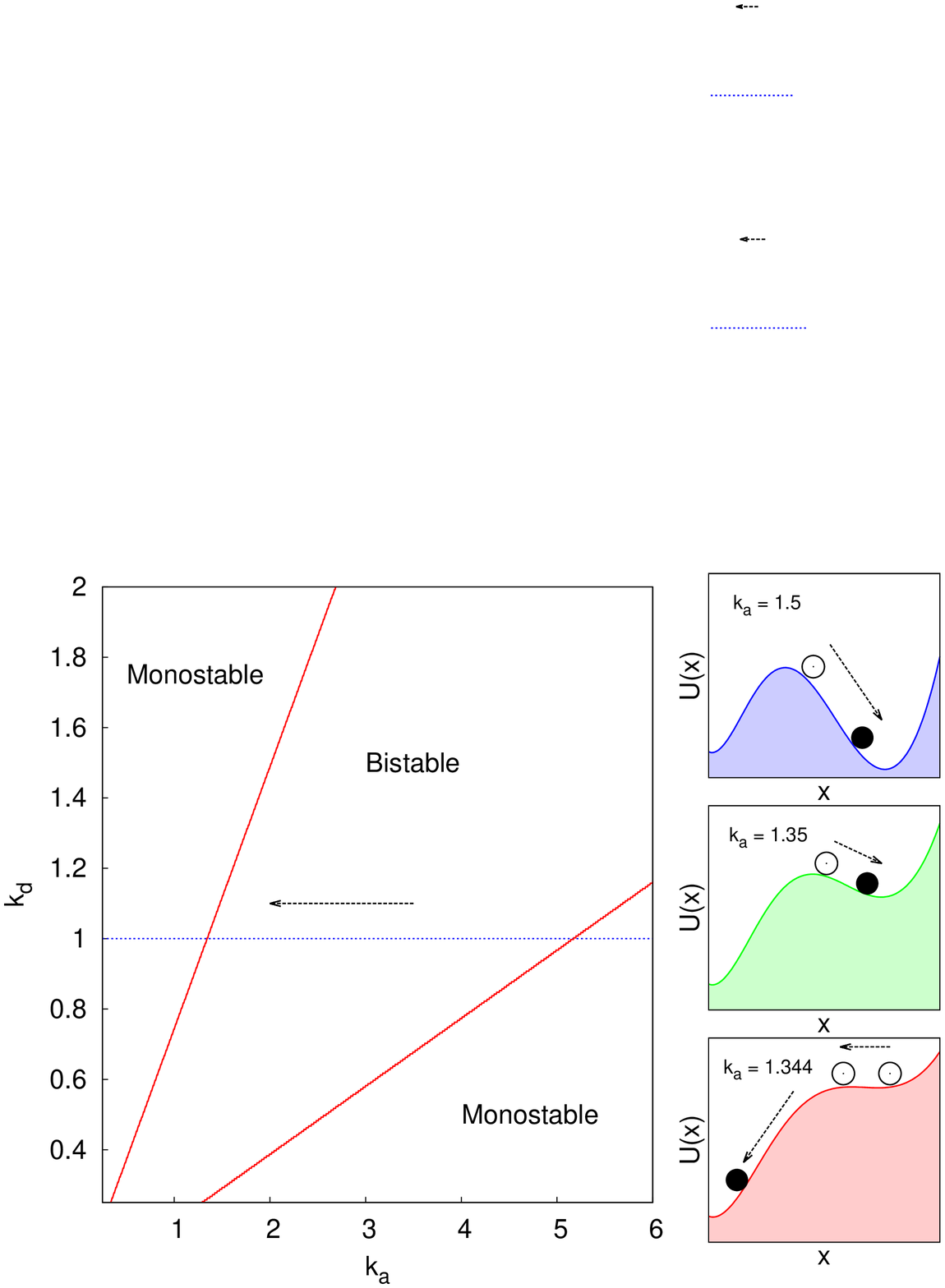}
\caption{The $k_{a}-k_{d}$ phase diagram of Model 1 showing a region
of bistability separating two regions of monostability. The stability
landscape $U(x)$ versus $x$ for three values of the bifurcation parameter
is also shown. As one moves toward the lower bifurcation point $k_{a}=1.344$,
indicated by an arrow in the region of bistability, the basin of attraction
corresponding to the high stable expression state becomes flatter.
At the bifurcation point itself, the return time $T_{R}$, associated
with the CSD, becomes zero.} 
\label{fig3}
\end{figure}

\begin{equation}
\frac{dx}{dt}=\frac{J_{1}k_{a}^{\prime}}{k_{a}^{\prime}+k_{d}}+\frac{J_{0}k_{d}}{k_{a}^{\prime}+k_{d}}-k_{p}x
\label{funrk}
\end{equation}
\noindent The steady state condition $\frac{dx}{dt}=0$ yields three solutions
in a specific parameter regime corresponding to two stable steady
states separated by an unstable steady state. The model dynamics exhibit
the fold-bifurcation of the type shown in Figure \ref{fig1}. The rate constants
$J_{0}$, $J_{1}$, and $k_{a}$ serve as the bifurcation parameters.
Figure \ref{fig2}(b) shows the genetic circuit (Model 2), proposed in Ref.
\cite{suel06}, as governing the dynamics of competence development via
excitability. The circuit contains the autoregulatory positive feedback
loop of Figure \ref{fig2}(a). In addition there is a negative feedback loop
in which the ComK protein represses the expression of the \textit{comS}
gene whereas the ComS protein inhibits the degradation
of ComK by the MecA-ClpP-ClpC complex. The repression of \textit{comS} by
ComK is however, not well established under wild-type
expression conditions \cite{samoilov06}. The dynamics of the model are
described in terms of the differential equations \cite{suel06}:

\begin{equation}
\frac{dK}{dt}=a_{k}+\frac{b_{k}\: K^{n}}{k_{0}^{n}+K^{n}}-\frac{K}{1+K+S}
\label{suel1}
\end{equation}
\begin{equation}
\frac{dS}{dt}=\frac{b_{s}}{1+(K/k_{1})^{p}}-\frac{S}{1+K+S}
\label{suel2}
\end{equation}
\noindent The variables $K$ and $S$ denote the concentrations of the ComK and ComS
proteins respectively, $a_{k}$ and $b_{k}$ represent the basal and
fully activated rates of ComK synthesis and $k_{0}$ is the ComK concentration
needed for 50\% activation. The Hill coefficients n and p are indicative
of the cooperativities of ComK autoactivation and ComS repression respectively.
The expression rate of ComS has maximal value $b_{s}$ and is half-maximal
at $K=k_{1}$. The non-linear degradation terms are a consequence
of the MecA complex-mediated degradation with a competitive inhibition
of the degradation by ComS.

\begin{figure}
\includegraphics[scale=0.6]{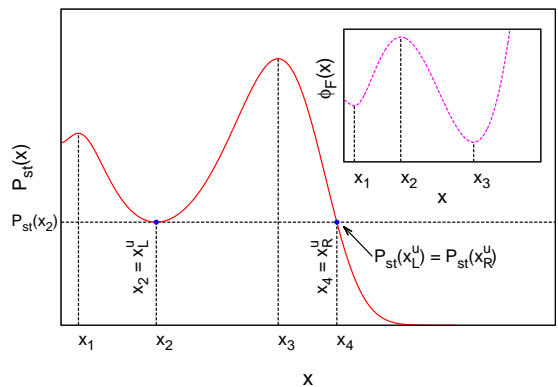}
\caption{The steady SSPD $P_{st}(x)$ of
protein levels in the region of bistability. The maxima of $P_{st}(x)$
at $x_{1}$ and $x_{3}$ represent low and high expression states respectively.
The points $x_{2}$ and $x_{4}$ denote the lower and upper cut-off
points of the probability distribution $P_{u}^{n}(x)$ corresponding
to the high expression state. The inset shows the stochastic potential
$\phi_{F}(x)$ versus $x$.} 
\label{fig4}
\end{figure}

\section{Early signatures of regime shifts}
\label{earlysig}

The deterministic dynamics of Model 1 are governed by the rate equation shown in Equation (\ref{funrk}). 
A simple stochastic version of the model has been studied in Ref.
\cite{karmakar07}. In this model, the only stochasticity considered is associated
with the random transitions of the gene between the inactive ($G$) and
active ($G^{*}$) states. The rest of the processes undergo deterministic
time evolution. The simple model yields bimodal protein distributions
in the steady state in certain parameter regions. In this paper, the
stochastic dynamics of the model are investigated using the formulations
based on the Langevin and Fokker-Planck (FP) equations. The steady
state analysis of a bistable system in these formalisms is described
in Refs. \cite{kampen92,gardiner83,risken84,bialek01,hasty00}. The one-variable Langevin equation (LE) including
both multiplicative and additive noise terms is given by 
\begin{equation}
\frac{dx}{dt}=f(x)+g_{1}(x)\varepsilon(t)+\Gamma(t)
\label{langevin}
\end{equation}
\noindent where $\varepsilon(t)$ (multiplicative noise) and $\Gamma(t)$ (additive
noise) are Gaussian white noises with mean zero and correlations:   
\begin{eqnarray}
\langle\varepsilon(t)\varepsilon(t^{\prime})\rangle&=&2d_{1}\delta(t-t^{\prime})\nonumber \\ 
\langle\Gamma(t)\Gamma(t^{\prime})\rangle&=&2d_{2}\delta(t-t^{\prime})\nonumber\\
\langle\varepsilon(t)\Gamma(t^{\prime})\rangle&=&\langle\Gamma(t)\varepsilon(t^{\prime})\rangle
=2\lambda_{d}\sqrt{d_{1}d_{2}}\delta(t-t^{\prime})
\label{correl}
\end{eqnarray}
In Equation (\ref{correl}), $d_{1}$ and $d_{2}$ denote the strengths of the noises
$\varepsilon(t)$ and $\Gamma(t)$ respectively and $\lambda_{d}$ is
the degree of correlation between them. The first term in Equation (\ref{langevin})
describes the deterministic dynamics. In the case of Model 1, $f(x)$
is given by the right hand side expression in Equation (\ref{funrk}). The FP equation
is a rate equation for the probability distribution $P(x,t)$, obtained
from the LE as \cite{gardiner83,fox88}:
\begin{equation}
\frac{\partial P(x,t)}{\partial t}=-\frac{\partial}{\partial x}[A(x)P(x,t)]+\frac{\partial^{2}}{\partial x^{2}}[B(x)P(x,t)]
\label{dist}
\end{equation}
where
\begin{equation}
A(x)=f(x)+d_{1}g_{1}(x)g_{1}^{\prime}(x)+\lambda_{d}\sqrt{d_{1}d_{2}}g_{1}^{\prime}(x)
\label{func1}
\end{equation}
and
\begin{equation}
B(x)=d_{1}[g_{1}(x)]^{2}+2\lambda_{d}\sqrt{d_{1}d_{2}}g_{1}(x)+d_{2}
\label{func2}
\end{equation}
\noindent The steady state probability distribution (SSPD) is given by \cite{gardiner83,risken84,bialek01} 
\begin{eqnarray}
P_{st}(x)&=&\frac{N}{B(x)}\exp\left[\int^{x}\frac{A(x)}{B(x)}dx\right]\nonumber \\
&=&\frac{N}{\{d_{1}[g_{1}(x)]^{2}+2\lambda_{d}\sqrt{d_{1}d_{2}}g_{1}(x)+d_{2}\}^{\frac{1}{2}}}\nonumber\\
&\times&\exp\left[\int^{x}\frac{f(x^{\prime})dx^{\prime}}
{d_{1}[g_{1}(x^{\prime})]^{2}+2\lambda_{d}\sqrt{d_{1}d_{2}}g_{1}(x^{\prime})+d_{2}}\right]\nonumber\\
\;
\label{big}
\end{eqnarray}
\noindent where $N$ is the normalization constant. Equation (\ref{big}) can be rewritten in
the form
\begin{equation}
P_{st}(x)=Ne^{-\phi_{F}(x)}
\label{dist2}
\end{equation}
\noindent with
\begin{eqnarray}
\phi_{F}(x)&=&\frac{1}{2}\ln\left[d_{1}[g_{1}(x)]^{2}+2\lambda_{d}\sqrt{d_{1}d_{2}}g_{1}(x)+d_{2}\right]\nonumber\\
&&-\int^{x}\frac{f(y)dy}{d_{1}\left[g_{1}(y)\right]^{2}+2\lambda_{d}\sqrt{d_{1}d_{2}}g_{1}(y)+d_{2}}
\label{stochpot}
\end{eqnarray} 
\noindent $\phi_{F}(x)$ defines the \textquotedblleft stochastic potential\textquotedblright$\,$ of the dynamics.
Once the SSPD is determined, quantities
like the variance and skewness, to be defined below, which provide
early signatures of regime shifts can be determined.

\begin{figure}
\includegraphics[scale=0.4]{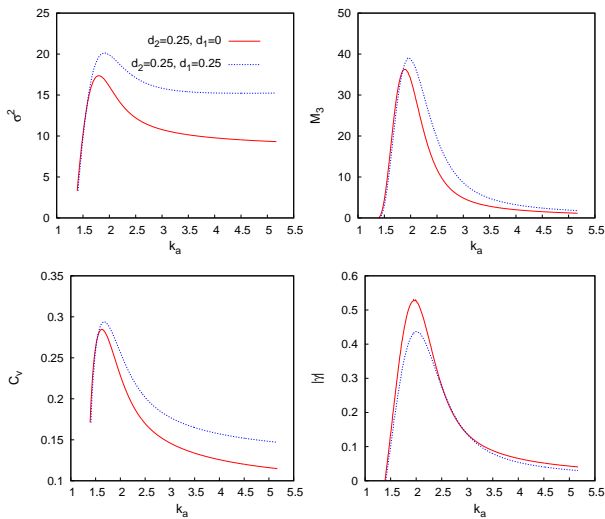}
\caption{ For Model 1, the plots of the variance $\sigma^{2}$,
the third moment $M_{3}$, the coefficient of variation $C_{v}$ and
the modulus $\left|\gamma\right|$ where $\gamma$ is the skewness of the normalized
SSPD $P_{u}^{n}(x)$ (Equation (\ref{norm2})) as
the bifurcation parameter $k_{a}$ is decreased towards the lower
bifurcation point. The solid lines correspond to the case when only
additive noise is present. The dotted lines are obtained when both
additive and multiplicative noise terms are present in the LE (Equation
(\ref{langevin})). The strengths of the multiplicative and additive noise are $d_{1}=0.25$
and $d_{2}=0.25$ respectively. The other parameter values are $k_{d}=1$,
$k_{s}=15$, $k_{p}=0.03$, $J_{0}=0.01$ and $J_{1}=1$.} 
\label{fig5}
\end{figure}

We next define the quantitative measures of the early signatures of
regime shifts. The first signature is that of the CSD which is a distinguishing
feature of the dynamics close to a bifurcation point \cite{vannes07,dakos12,wissel84}.
This involves a progressively larger relaxation time, as the bifurcation
point is approached, to an attractor of the dynamics (say, a stable
steady state) after being weakly perturbed from it. The physical origin
of this phenomenon can be understood in terms of the stability landscape
$U(x)$. The rate equation governing the dynamics of Model 1 (Equation (\ref{funrk}))
can be written as $\frac{dx}{dt}=f(x)=-\frac{\partial U(x)}{\partial x}$.
Figure \ref{fig3}(a) shows the phase diagram of the model in the $k_{a}-k_{d}$
plane with a region of bistability separating two regions of monostability.
The other parameter values are 
$J_{1}=1$, $J_{0}=0.01$, $k_{p}=0.03$ in appropriate units. Figure
\ref{fig3}(b) shows the three stability landscapes at three successively decreasing values
of $k_{a}$ as one approaches the bifurcation
point in the direction of the arrow from within the region of bistability.
At point $1$ ($k_{a}=1.5$), the states corresponding to the minima of $U(x)$ represent
the stable steady states. The location of the \textquotedblleft ball\textquotedblright$\,$ represents
the state of the dynamical system. As one progresses from point 1
to point 2 ($k_{a}=1.35$), the steady state with high value of $x$ becomes less stable.
The associated basin of attraction becomes flatter with reduced size
so that it takes a longer time for a perturbed state (ball shifted
from the minimum position) to relax back to the stable steady state.
The relaxation time diverges at the bifurcation point where the stable
steady state is on the verge of losing stability (point 3, $k_{a}=1.344$). A weak
perturbation pushes the ball to the minimum with low value of $x$,
i.e., the system does not relax back to the high $x$ state. One can
define a return time $T_{R}$ which provides a measure of the time
taken by the dynamical system to regain a stable steady state after
being weakly perturbed from it. Let $x_{st}$ be the stable steady
state value of $x$ and $\eta(t)=x(t)-x_{st}$ be the small deviation
from the steady state value under weak perturbation. The deterministic
rate equation is given by $\frac{dx}{dt}=f(x).$ On Taylor expanding
$f(x)$ around $x=x_{st}$ and keeping terms upto the order of $\eta(t),$
one obtains
\begin{equation}
\frac{d\eta}{dt}=\lambda\eta(t),\;\lambda=\left.\frac{df(x)}{dx}\right|_{x=x_{st}}
\label{rateeq}
\end{equation}
The solution of the Equation (\ref{rateeq}) is
\begin{equation}
\eta(t)=\eta(0)e^{\lambda t}
\label{eta}
\end{equation}
\noindent where $\eta(0)$ is the initial value of $\eta(t)$ at time $t=0$.
The sign of the parameter $\lambda$ indicates the stability of the
steady state. If $\lambda$ is $<0$ ($>0$), the steady state is
stable (unstable). We designate $\lambda$ as the stability parameter.
It is a well-known result from dynamical systems theory that at a
bifurcation point, the stability parameter $\lambda$, associated
with the steady state losing stability, becomes zero \cite{stogatz94,vannes07,dakos12,wissel84}.
In the case of Model 1, one can check that $\lambda=0$ at the two
bifurcation points. The return time $T_{R}=\frac{1}{\left|\lambda\right|}$ thus
diverges at a bifurcation point. If the dynamical system is described
by more than one variable, the stability of a steady state is determined
by the eigenvalues of the Jacobian matrix governing the dynamics of
the perturbed system \cite{stogatz94}. Let $\lambda_{max}$ be the real part
of the dominant eigenvalue of the Jacobian matrix (for stability,
the real parts of all the $\lambda$'s should
be negative). The dominant eigenvalue is the one with the largest
real part. The return time $T_{R}$ is given by $T_{R}=\frac{1}{\left|\lambda_{max}\right|}$.
Examples of the experimental observations of the CSD include the transition
from the $G2$ growth phase to the mitotic phase of the eukaryotic
cell division cycle \cite{hasty00}, a direct observation of the CSD in a
laboratory population of budding yeast before population collapse
occurs at a critical experimental condition \cite{dai12} and the demonstration
of the CSD in a population of cyanobacteria \cite{veraart12}.

\begin{figure}
\includegraphics[scale=0.6]{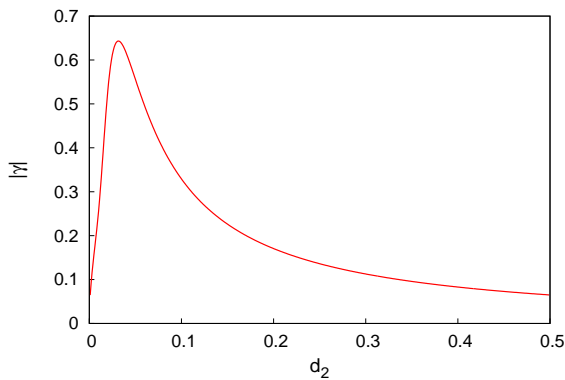}
\caption{ For Model 1, the plot of $\mid\gamma\mid,$ where $\gamma$
is the skewness of the normalized steady SSPD,
versus the strength of additive noise $d_{2}$ for the parameter values
$k_{a}=1.5$, $k_{d}=1.5$, $k_{p}=0.03$, $J_{0}=0.01$, $J_{1}=1$,
and $d_{1}=0$ (no multiplicative noise). The value $k_{a}=1.5$ falls
in the region of bistability.} 
\label{fig6}
\end{figure}

In the presence of noise, the variance $\sigma^{2}$ is determined
from the fluctuation-dissipation (FD) relation (Equation (\ref{FD})) in
Appendix A. In the case of a one-variable model (Model 1), the matrix
$\mathbf{A}$ consists of a single element $\lambda$, the stability parameter
defined in Equation (\ref{rateeq}). The steady state covariance matrix $\mathbf{C}$ reduces
to a single element, the variance $\sigma^{2}$, which is determined
as 
\begin{equation}
\sigma^{2}=\frac{D}{2\mid\lambda\mid}
\end{equation}
with $\lambda$ negative. Also, the lag-1 autocorrelation in the stationary
state (Equations (\ref{acorrel}) and (\ref{autolag}) with $\tau=1$) is given by 
\begin{equation}
\rho(1)=e^{\lambda}
\end{equation}
The variance $\sigma^{2}$ diverges and the lag-1 autocorrelation
$\rho(1)$ reaches its maximum value at the bifurcation point since
the stability parameter $\lambda$ is zero at this point. Rising variance and autocorrelation thus provide
early signatures of impending regime shifts. The CSD close to the
bifurcation point implies that the system's intrinsic rates of change
are decreased so that the state of the system at time $t$ resembles
closely the state at time $t-1$. This increased memory is measured
by the lag-1 autocorrelation function. Also, because of a flatter
basin of attraction close the transition point, a given perturbation
brings about a greater shift in the system's state, i.e., an increasing
variance.

\begin{figure}
\includegraphics[scale=0.45]{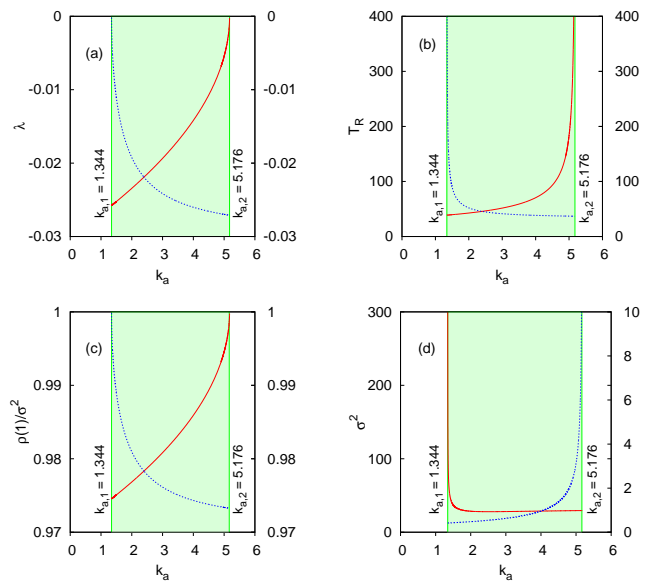}
\caption{ For Model 1, the variation of the stability parameter $\lambda$
(Equation (\ref{rateeq})), the return time $T_{R}$, the lag-1 autocorrelation $\rho_{\eta}(1)$
and the variance $\sigma^{2}$ (Equations (17) and (18)) as a function
of the bifurcation parameter. The plots exhibit characteristic features
at the bifurcation points $k_{a,1}=1.344$ and $k_{a,2}=5.176$. The
parameter values are $k_{d}=1$, $k_{s}=15$, $k_{p}=0.03$, $J_{0}=0.01$,
$J_{1}=1$ and $\sigma_{d}=0.25$.} 
\label{fig7}
\end{figure}

\begin{figure}
\includegraphics[scale=0.4]{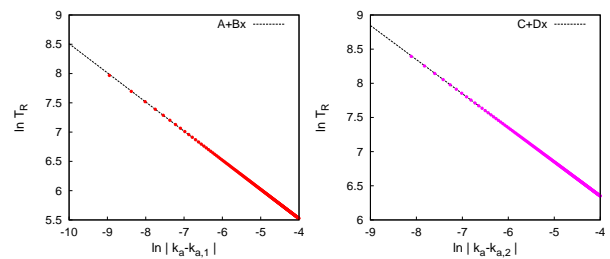}
\caption{The computed data points in the $\ln T_{R}$ versus (a) $\ln \left|k_{a}-k_{a,1}\right|$ and (b) 
 $\ln \left|k_{a}-k_{a,2}\right|$ plots are fitted by straight lines (a) $y=A+Bx$ and (b) $y=C+Dx$ respectively. The 
 parameter values are the same as in the case of Figure \ref{fig7}. The values of $A$ and $C$ are 
 $A=3.53401\pm 0.00068$ and $C=4.35482\pm 0.00049$. The exponents have values very close to $-\frac{1}{2}$ 
 ($B=-0.49752\pm 0.00013$ and $D=-0.49901\pm 0.00099$).} 
\label{fig8}
\end{figure}

Two other early signatures of a regime shift which are not related
to the CSD are skewness and flickering \cite{scheffer091,guttal08}. The skewness $\gamma$
of a probability distribution $P(x)$ is a dimensionless measure of its
degree of asymmetry and is defined as 
\begin{equation}
\gamma=\frac{\int(x-\mu)^{3}P(x)dx}{\sigma^{3}}
\label{skewness}
\end{equation}
where $\mu$ and $\sigma$ are respectively the mean and standard
deviation of $P(x)$. The variance $\sigma^{2}$, the coefficient of
variation $C_{v}$ and the third moment $M_{3}$ of $P(x)$ are expressed
as  
\begin{eqnarray}
\sigma^{2}&=&\int(x-\mu)^{2}P(x)dx\nonumber \\
M_{3}&=&\int(x-\mu)^{3}P(x)dx\nonumber\\
C_{v}&=&\frac{\sigma}{\mu}
\label{moments}
\end{eqnarray}   
\noindent The skewness of a probability distribution increases as the bifurcation
point is approached. This is because of the asymmetry in fluctuations
with a system exhibiting greater amplitude deviations in the direction
of the state it is fated to switch to than in the other direction.
The phenomenon of flickering is observed in the region of bistability
with the system switching back and forth between the two attractor
states before the bifurcation point is reached. We propose a quantitative
measure of flickering which serves as an early signature of regime
shift. In the region of bistability, the stability landscape has two
minima corresponding to the two stable steady states. Noise-induced
transitions take the system from one valley to the other. The mean
first passage time (MFPT) refers to the average first exit time from
a valley \cite{kampen92,gardiner83}. Let $T_{1}$ and $T_{2},$ be the MFPTs for
the exits respectively from valley 1 and valley 2. The times are indicative
of the amount of flickering present in the system. The MFPT $T_{2}$
becomes zero at the bifurcation point where the steady state 2 loses
stability. The ratio $r=\frac{T_{1}}{T_{2}}$ measures the asymmetry
in the exit times and diverges as the bifurcation point, at which
$T_{2}\rightarrow0$ is approached. The quantity $r$ thus provides
an early signature of regime shift.

\begin{figure}
\includegraphics[scale=0.6]{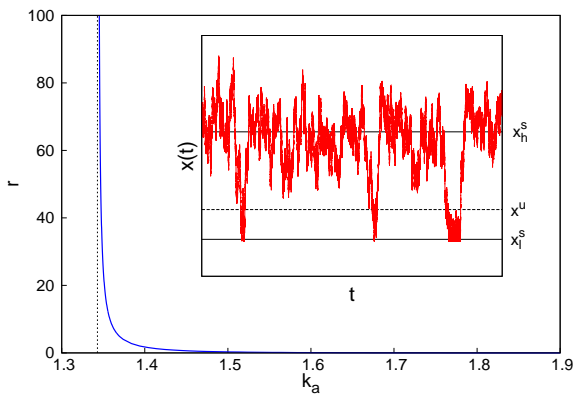}
\caption{ For Model 1, the plot of $r=\frac{T_{1}}{T_{2}}$ versus
the bifurcation parameter $k_{a}$ where $T_{1}$ and $T_{2}$ are
the MFPTs. The ratio $r$ is seen to diverge at the lower bifurcation
point. Inset shows the time-series data of $x(t)$ versus $t$ obtained by
solving the LE. The straight lines are drawn at the steady state points
$x_{h}^{s}$, $x^{u},$ $x_{l}^{s}$ obtained by solving the deterministic
rate equation, Equation (\ref{funrk}). The parameter values are $k_{d}=1$, $k_{s}=15$,
$k_{p}=0.03$, $J_{0}=0.01$, $J_{1}=1$ and $d_{2}=0.25$. } 
\label{fig9}
\end{figure}

\section{Results on Early Signatures}
\label{results}

\subsection{Model 1 (one variable)}
We first present the results on early signatures in the case of Model
1. In the region of bistability, the SSPD
$P_{st}(x)$ is bimodal with two distinct peaks. $P_{st}(x)$ may
be obtained via a solution of the FP equation (Equation (\ref{big})) or obtained
by a numerical solution of the LE (Equation(\ref{langevin})). Figure \ref{fig4} shows an example
of a bimodal distribution with two distinct peaks at $x=x_{1}$ and
$x_{3}.$ The inset of the figure shows the corresponding stochastic
potential profile (Equation (\ref{stochpot})) with $x_{2}$ denoting the location of
the hill and $x_{1},$ $x_{3}$ denoting the minima of the left and
right valley respectively. We find the normalized probability distributions,
$P_{l}^{n}(x)$ and $P_{u}^{n}(x)$, corresponding to the low and
high expression states respectively, using the following procedure:

\begin{enumerate} 
\item From the stochastic potential profile $\phi_{F}(x)$, determine
the position of the hill at $x_{2}$ and compute $P_{st}(x_{2})$. The
point $x_{2}$ corresponds to the minimum of the probability distribution
between its two peaks at $x_{1}$and $x_{3}.$ The point $x_{2}$
is chosen to be the left cut-off point, $x_{L}^{u}$, for $P_{u}^{n}(x)$. 
The right cut-off point, $x_{R}^{u}$, of $P_{u}^{n}(x)$ is obtained
as a solution of the equation 
\begin{equation}
P_{st}(x)=P_{st}(x_{2}^{u}),\; x>x_{3}
\label{cut1}
\end{equation}
\item If $P_{st}(0)<P_{st}(x_{2})$, then the lower-cutoff point, $x_{L}^{l}$,
of $P_{l}^{n}(x)$ is determined from the solution of the equation 
\begin{equation}
P_{st}(x)=P_{st}(x_{R}^{l}),\; x<x_{1}
\label{cut2}
\end{equation}
\noindent where $x_{R}^{l}$, the right cut-off point of $P_{l}^{n}(x)$, is
given by $x_{R}^{l}=x_{L}^{u}$. If $P_{st}(0)>P_{st}(x_{2})$, $x_{L}^{l}=0$.
\end{enumerate}
\noindent Other cut-off procedures, e.g., that in Ref. \cite{guttal08}, have been proposed to isolate the low and high
expression probability distributions but the general results on the early signatures are qualitatively similar.    

In the case of Model 1, we first consider only an additive noise in
the LE (Equation (\ref{langevin}), $g_{1}(x)=0$). The additive noise $\Gamma(t)$ represents
noise arising from an external perturbative influence or originating from 
some missing information because of rate equation approximations \cite{hasty00}.
 The SSPD,
$P_{st}(x)$, in the region of bistability is obtained from Equation (\ref{big})
putting $g_{1}(x)$, $g_{1}(x^{\prime})=0$. Following the prescription already given, one
determines the cut-off points of the distribution $P_{l}^{n}(x)$.
The normalized distributions are obtained from  
\begin{equation}
P_{l}^{n}(x)=\frac{P_{st}^{\prime}(x)}{\int_{x_{L}^{l}}^{x_{R}^{l}}P_{st}(x)dx}\;\; \mbox{for}\;\; x_{L}^{l}<x<x_{R}^{l}
\label{norm1}
\end{equation} 
\begin{equation}
P_{u}^{n}(x)=\frac{P_{st}^{\prime}(x)}{\int_{x_{L}^{u}}^{x_{R}^{u}}P_{st}(x)dx}\;\; \mbox{for}\;\; x_{L}^{u}<x<x_{R}^{u}
\label{norm2}
\end{equation}
\noindent In Equations (\ref{norm1}) and (\ref{norm2}), $P_{st}^{'}(x)$ is not normalized. With a
knowledge of the normalized probability distributions $P_{l}^{n}(x)$ and
$P_{u}^{n}(x)$, one can compute the skewness $\gamma$, variance
$\sigma^{2}$, third moment $M_{3}$ and the coefficient of variation
$C_{v}$ using Equations (\ref{skewness}) and (\ref{moments}). Figure \ref{fig5} shows plots of these quantities
(solid lines) versus the bifurcation parameter $k_{a}$ for the probability
distribution $P_{u}^{n}(x)$, i.e., considering the system to be in
the high expression state. The sudden regime shift in the deterministic
case occurs at the lower bifurcation point $k_{a,1}=1.334$. The parameter
values used for the computation are: $k_{d}=1$, $k_{s}=15$, $k_{p}=0.03$,
$J_{0}=0.01$ and $J_{1}=1$. The strength $d_{2}$ of the additive
noise is $d_{2}=0.25$. One finds that all the four quantities $\left|\gamma\right|$,
$\sigma^{2}$, $M_{3}$ and $C_{v}$ increase as the lower bifurcation
point is approached thus providing early signatures of a regime shift.
The quantities, however, reach their maxima before the deterministic
bifurcation point is reached and then start decreasing. The quantities,
though providing early signatures, cannot provide knowledge of the
bifurcation point. We next include an additional multiplicative noise
term in the LE (Equation (\ref{langevin})) with  
\begin{equation}
g_{1}(x)=\frac{k_{a}^{\prime}}{k_{a}^{\prime}+k_{d}}
\end{equation}
\noindent The multiplicative noise is associated with the protein synthesis rate
constant $J_{1}$ in the active state, i.e., $J_{1}\rightarrow J_{1}+\varepsilon(t)$. 
The origin of multiplicative noise lies in the fact that the rate constants are expected 
to fluctuate in time due to the inherently stochastic nature of gene expression as well as 
due to stochastic influences like fluctuations in the number of regulatory molecules and RNA polymerases. 
With both the additive and multiplicative noise terms present, the
SSPDs $P_{l}^{n}(x)$ and $P_{u}^{n}(x)$
are computed following the procedure already described. The dotted
curves in Figure \ref{fig5} show the variations of $\left|\gamma\right|$, $\sigma^{2}$,
$M_{3}$ and $C_{v}$ , associated with the probability distribution
$P_{u}^{n}(x)$, as a function of the bifurcation parameter $k_{a}$.
The parameter values are the same as before with $d_{1}$, the strength
of the multiplicative noise term, having the value $d_{1}=0.25$.
The cross-correlation coefficient $\lambda_{d}$ in Equation (\ref{correl}) is taken
to be zero. Figure \ref{fig6} shows how the skewness of $P_{u}^{n}(x)$ changes
as a function of the additive noise strength $d_{2}$ for the parameter
values $k_{a}=1.5$, $k_{d}=1$, $k_{s}=15$, $k_{p}=0.03$, $J_{0}=0.01$,
$J_{1}=1$ and $d_{1}=0$ (no multiplicative noise). The value $k_{a}=1.5$
is greater than the value of $k_{a,1}=1.344$. The rising skewness
is thus a signature of noise-induced regime shift. In the case of
the SSPD $P_{l}^{n}(x)$, one obtains
early signatures of the upper bifurcation point $k_{a,2}=5.176$ similar
to the ones shown in Figures \ref{fig5} and \ref{fig6} in the case of $P_{u}^{n}(x)$,
though the quantitative measures exhibit less prominent variation. We find that 
the early signatures of regime shifts are obtained in both the cases, 
(i) only additive noise is present and (ii) additive as well as multiplicative 
types of noise are present.

\begin{figure}
\includegraphics[scale=0.6]{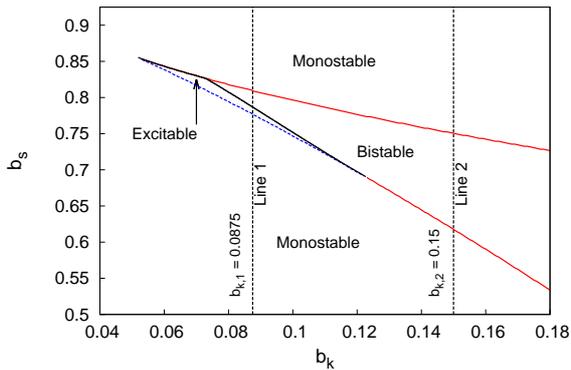}
\caption{ The phase diagram of Model 2 in the $b_{k}-b_{s}$ plane.
It shows two regions of monostability, one region of bistability
and one region of excitability. The boundaries
shown as solid lines correspond to saddle-node (SN) bifurcations.
Lines 1 and 2 mark the values of the parameter $b_{s}$ for which
computations are carried out. The other parameter values are the same
as in Ref. \cite{suel06}, namely, $a_{k}=0.004$, $k_{0}=0.2$, $k_{1}=0.222,$
$n=2$ and $p=5$.} 
\label{fig10}
\end{figure}

Considering Model 1, we next calculate the stability parameter $\lambda$
(Equation (\ref{rateeq})), with $f(x)$ given by the expression on the r.h.s. of Equation (\ref{funrk}), 
in the region of bistability. Knowing $\lambda$, the return
time $T_{R}(=\frac{1}{\left|\lambda\right|})$ as well as the lag-1 autocorrelation
$\rho(1)$ and the variance $\sigma^{2}$ (Equations (17)
and (18)) are also determined. Figure \ref{fig7} shows the variation of $\lambda$,
$T_{R}$, $\rho(1)$ and $\sigma^{2}$ as a function of the
bifurcation parameter $k_{a}$ and for the parameter values $k_{d}=1$,
$k_{s}=15$, $k_{p}=0.03$, $J_{0}=0.01$, $J_{1}=1$ and $\sigma_{d}=0.25$.
The stability parameter $\lambda$ becomes zero at the lower ($k_{a,1}=1.344)$
and the upper $(k_{a,2}=5.176)$ bifurcation points as expected. The
associated return time $T_{R}$ diverges at the bifurcation points,
a characteristic feature of the CSD. The lag-1 autocorrelation $\rho(1)$
reaches the maximum value and the variance $\sigma^{2}$ diverges
at the bifurcation points. These quantities are good indicators of
regime shifts and carry distinct signatures of the bifurcation points. The return time $T_{R}$ is 
known to satisfy a general scaling law $T_{R}\sim \left|B-B_{i}\right|^{-\frac{1}{2}}$ where $B$ and $B_{i}$
stand for the bifurcation parameter and point respectively \cite{wissel84}. In Figure \ref{fig8}, the scaling 
is demonstrated for the bifurcation parameter $k_{a}$ and the bifurcation points $k_{a,1}=1.344$ and $k_{a,2}=5.176$. 
The exponent is very close to $-\frac{1}{2}$ in each case.

Figure \ref{fig9} shows the variation of $r={T_{1}}/{T_{2}}$ versus
the bifurcation parameter $k_{a}$. The ratio is seen to diverge at
the lower bifurcation point $k_{a,1}=1.344$. The parameter values
are $k_{d}=1$, $k_{s}=15$, $k_{p}=0.03$, $J_{0}=0.01$, $J_{1}=1$
and $d_{2}=0.25$ (only additive noise is considered). The time-series
data shown in the inset of Figure \ref{fig9} is obtained via numerical solution
of the LE (Equation (\ref{langevin})) using the algorithm described in Ref. \cite{fox88}.
In the limit of large times, the steady state is assumed to be reached.
The solid lines in the inset mark the stable expression states $x_{l}^{s}$
and $x_{h}^{s}$ and the dotted line corresponds to the unstable steady
state $x^{u}$. The values are obtained from a solution of the deterministic
rate equation. The MFPTs $T_{1}$ and $T_{2}$ are computed using
the method outlined in Ref. \cite{ghosh12}. Let us consider a bistable potential
with the stable steady states at $x_{1}$ and $x_{3}$ ($x_{1}<x_{3}$)
which are separated by an unstable steady state at $x_{2}$, termed
the barrier state (Figure \ref{fig4}). The MFPT, $T(x;a,b)$ is
the average time of the first exit from the interval $(a,b)$ and satisfies
the equation \cite{gardiner83,gillespie92} 
\begin{equation}
-1=A(x)\frac{dT(x)}{dx}+\frac{1}{2}B(x)\frac{d^{2}T(x)}{dx^{2}}
\label{mfpt}
\end{equation}
\noindent where $A(x)$ and $B(x)$ appear in the associated FP equation (Equation (\ref{dist})). 
The MFPT $T(x_{1})(=T_{1})$ for exit from the basin of attraction
of the stable steady state at $x_{1}$ is obtained as a solution of
Equation (\ref{mfpt}) with the interval $(a,b)=(0,x_{2})$ and boundary conditions
given by 
\begin{equation}
T^{\prime}(a;a,b)=0\;\; \mbox{and}\;\; T(b;a,b)=0
\label{bc1}
\end{equation}
\noindent The prime denotes differentiation with respect to $x$, with reflecting
and absorbing boundary conditions prevailing at $a$ and $b$ respectively
\cite{gardiner83,gillespie92}. Following the same procedure, the MFPT $T(x_{3})(=T_{2})$
for exit from the basin with the stable steady state at $x_{3}$ can
be calculated from Equation (\ref{mfpt}). The interval now is $(a,b)=(x_{2},\infty)$
with $x_{2}$ and $\infty$ serving as absorbing and reflecting boundary
points respectively, i.e., 
\begin{equation}
T(a;a,b)=0\;\;\mbox{and}\;\;T^{\prime}(b;a,b)=0
\label{bc2}
\end{equation}

\begin{figure}
\includegraphics[scale=0.45]{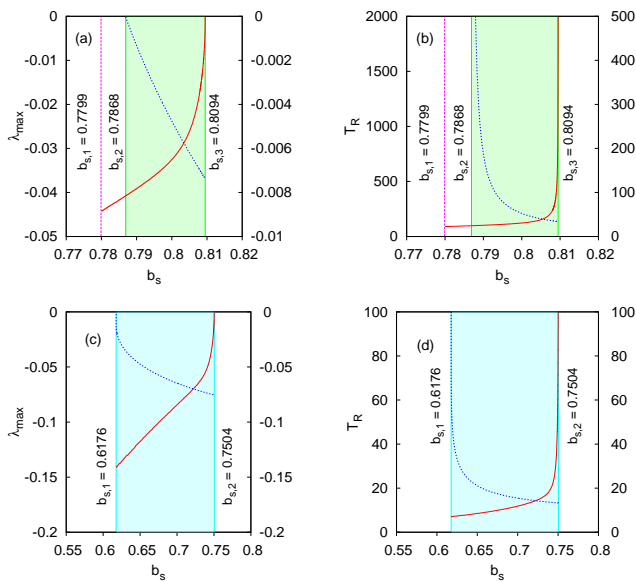}
\caption{For Model 2, the variation of $\lambda_{max}$ and $T_{R}$
as a function of $b_{s}$. (a), (b): the computations are carried
out along the line $b_{k,1}=0.0875.$ (c), (d): the computations are
carried out along the line $b_{k,2}=0.15.$ The parameter values are
the same as in the case of Figure \ref{fig10}.} 
\label{fig11}
\end{figure}

\subsection{Model 2 (two variables)}
We next consider Model 2 the dynamics of which are governed by the
set of two equations (\ref{suel1}) and (\ref{suel2}). In Ref. \cite{suel06}, the phase diagram
of the model in the $b_{k}-b_{s}$ plane has been determined which
has four different regions: (i) monostability with only one fixed
point, (ii) bistability with three fixed points, two stable and one
unstable, (iii) excitability involving three fixed points only one
of which, corresponding to low ComK value, is stable. The competent
fixed point (high ComK level) has the characteristic of an unstable
spiral and the mid-ComK fixed point is a saddle point, (iv) only one
fixed point exists which is unstable. The system exhibits limit cycle
oscillations between the mid-ComK and high-ComK levels. For the purpose
of our study, a part of the $b_{k}-b_{s}$
phase diagram has been recomputed and shown in Figure \ref{fig10}. The diagram
shows three different regions: monostable, bistable and excitable.
The monostable region is again of two types: monostable low (low ComK
level) and monostable high (high ComK level). The boundaries between
the regions depicted by solid lines correspond to the saddle-node
(SN) bifurcation \cite{stogatz94}. Two vertical lines 1 and 2 are drawn in
the phase diagram at the points $b_{k,1}=0.0875$ and $b_{k,2}=0.15$
respectively. Line 1 intersects the phase boundaries at the three points
$b_{s,1}=0.7799$, $b_{s,2}=0.7868$ and $b_{s,3}=0.8094$. Line 2
has two points of intersection: $b_{s,1}=0.6175$ and $b_{s,2}=0.7504$.
Figure \ref{fig11} shows the plots of $\lambda_{max}$, the real part of the
dominant eigenvalue of the Jacobian matrix computed from Equations (\ref{suel1})
and (\ref{suel2}), and the return time $T_{R}=\frac{1}{\left|\lambda_{max}\right|}$
versus $b_{s}$ along line 1 ((a) and (b)) and along line 2 ((c) and
(d)). For a specific steady state, the Jacobian matrix $\mathbf{J}$ has the following
structure \cite{stogatz94}: 
\begin{equation}
\mathbf{J}=\left(
\begin{array}{cc}
f_{11} & f_{12}\\
f_{21} & f_{22}
\end{array}
\right)_{SS}
\label{jacobi}
\end{equation}
\noindent The suffix $SS$ stands for steady state, i.e., the matrix elements are to be computed at
the fixed point ($x^{*},y^{*})$. A matrix element $f_{ij}=\frac{\partial f_{i}}{\partial x_{j}}$
($i=1,2$, $j=1,2$) where
\begin{equation}
f_{1}(x_{1},x_{2})=a_{k}+\frac{b_{k}\,x_{1}^{n}}{k_{0}^{n}+x_{1}^{n}}-\frac{x_{1}}{1+x_{1}+x_{2}}
\label{sfun1}
\end{equation}
\begin{equation}
f_{2}(x_{1},x_{2})=\frac{b_{s}}{1+(x_{1}/k_{1})^{p}}-\frac{x_{2}}{1+x_{1}+x_{2}}
\label{sfun2}
\end{equation}
\noindent with $x_{1}=K$, $x_{2}=S$, $n=2$ and $p=5$. The parameter values
are the same as in Ref. \cite{suel06}: $a_{k}=0.005$, $k_{0}=0.2$, and
$k_{1}=0.222$. The eigenvalue of $\mathbf{J}$ are $\lambda_{1}$ and $\lambda_{2}$
and $\lambda_{max}$ is the real part of the dominant eigenvalue.
In Figure \ref{fig11}, one notes that $\lambda_{max}$ becomes zero at the
SN bifurcation points as expected and the corresponding return time
$T_{R}$ diverges at a bifurcation point. The shaded regions
in the Figure denote the regions of bistability in which the two stable
steady states correspond to low and high ComK levels respectively.
The level of ComS is anticorrelated with that of ComK. The solid (dotted)
lines in Figure \ref{fig11} are associated with the stable steady states representing
low (high) ComK levels. Along line 2 and at the bifurcation point
$b_{s,1}=0.6176$ (Figures \ref{fig11}(c) and (d)) the stable steady state
corresponding to the high ComK level loses stability ($\lambda_{max}=0$,
$T_{R}$ diverges). At the boundary point $b_{s,2}=0.7504$, the state
representing the low ComK level loses stability. A steady state is
stable if the real parts of both $\lambda_{1}$ and $\lambda_{2}$
are negative. At the point $b_{s,1}=0.7799$ along line 1 (Figures
\ref{fig11}(a) and (b)), there is no loss of stability of a steady state and
hence no CSD with diverging $T_{R}$ is observed at this point. The
low ComK level of the monostable region continues to remain stable
as one enters the region of excitability at the point $b_{s,1}$ along
line 1. At the point $b_{s,2}=0.7868,$ the system enters the region
of bistability from a region of excitability. The high ComK stable
steady state loses its stability at this point with $\lambda_{max}=0$
and a divergent $T_{R}$ (dotted branch in Figures \ref{fig11}(a) and (b)).
The low ComK state loses stability at the point $b_{s,3}=0.8094$
when the system passes from a region of bistability to a region of
monostable high ComK level. The eigenvalue $\lambda_{max}=0$ and
$T_{R}$ diverges at this point. The main point to note from the results
is that there is no CSD and diverging return time $T_{R}$ at the
transition point $b_{s,1}$ along line 1 between the regions of monostability
and excitability. On the other hand, at the point $b_{s,1}$ along line 2, 
separating the regions of monostability and bistability, one of the expression states, 
namely, the high ComK state, loses stability with a divergent $T_{R}$.

We next consider the time evolution of the ComK-ComS system to be stochastic in nature.
As in the case of Model 1, the FD relation (Equation (\ref{FD})) can be
used to calculate the stationary state variances of the fluctuations around
the steady state and also the lag-1 autocorrelation function $\rho_{11}(\tau=1)$
(Equations (\ref{acorrel}) and (\ref{autolag})). The Jacobian matrix $A$ has the form shown
in Equation (\ref{jacobi}) with the elements calculated from Equations (\ref{sfun1})
and (\ref{sfun2}) using the relationship $f_{ij}=\frac{\partial f_{i}}{\partial x_{j}}$
$(i=1,2,\; j=1,2)$. The stoichiometric matrix is given by 

\begin{equation}
\mathbf{S}=\left(
\begin{array}{cccc}
1 & -1 & 0 & 0\\
0 & 0 & 1 & -1
\end{array}\right)
\label{stoichio}
\end{equation}
The first and second rows of \textbf{$\mathbf{S}$} correspond to
ComK and ComS molecular numbers respectively. The \textquotedblleft elementary
complex reactions\textquotedblright \cite{kampen92,elf} considered for the
computations are the composite reactions (Equations (\ref{suel1}) and (\ref{suel2}) with
$x_{1}=K$ and $x_{2}=S$ ),

\begin{equation}
\begin{array}{c}
X_{1}\overset{a_{k}+\frac{b_{k}\,x_{1}^{n}}{k_{0}^{n}+x_{1}^{n}}}{\longrightarrow}X_{1}+1\\
X_{1}\overset{\frac{x_{1}}{1+x_{1}+x_{2}}}{\longrightarrow}X_{1}-1\\
X_{2}\overset{\frac{b_{s}}{1+(\frac{x_{1}}{k_{1}})^{p}}}{\longrightarrow}X_{2}+1\\
X_{2}\overset{\frac{x_{2}}{1+x_{1}+x_{2}}}{\longrightarrow}X_{2}-1
\end{array}
\label{comporeact}
\end{equation}

The reaction propensity vector (Equation (\ref{arate2})) is given by

\begin{equation}
\mathbf{f}(\mathbf{x})=
\left(
\begin{array}{cccc}
a_{k}+\frac{b_{k}x_{1}^{n}}{k_{0}^{n}+x_{1}^{n}} & 0 & 0 & 0\\
0 & \frac{x_{1}}{1+x_{1}+x_{2}} & 0 & 0\\
0 & 0 & \frac{b_{s}}{1+(\frac{x_{1}}{k_{1}})^{p}} & 0\\
0 & 0 & 0 & \frac{x_{2}}{1+x_{1}+x_{2}}
\end{array}\right)
\label{propens}
\end{equation}

\begin{figure}
\includegraphics[scale=0.45]{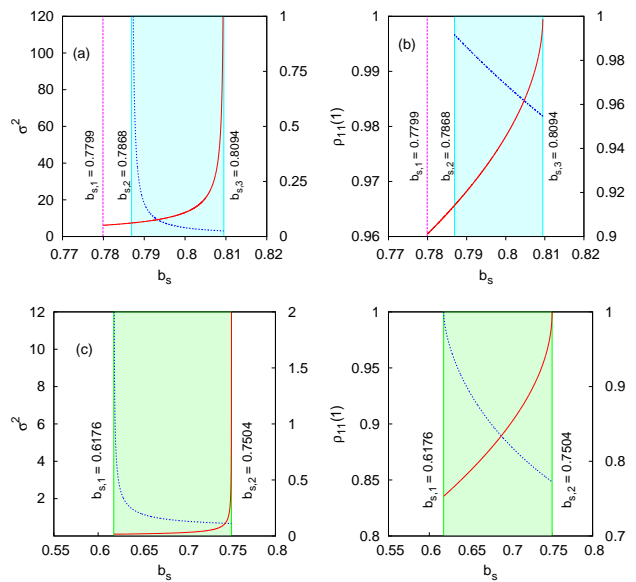}
\caption{For Model 2, the variation of $\sigma^{2}$ and $\rho_{11}(1)$
as a function of $b_{s}$. (a), (b): the computations are carried
out along the line $b_{k,1}=0.0875.$ (c), (d): the computations are
carried out along the line $b_{k,2}=0.15.$ The parameter values are
the same as in the case of Figure \ref{fig10}.} 
\label{fig12}
\end{figure}

With knowledge of the stoichiometric matrix $\mathbf{S}$ and the
reaction propensity vector $\mathbf{f(x)}$, the diffusion
matrix $D$ in the FD relation can be calculated using Equation (\ref{dmat}).
Substituting the computed $\mathbf{A}$ and $\mathbf{D}$ matrices in the FD relation, the
variances and the covariances are determined. Similarly, with the
help of Equations (\ref{acorrel}) and (\ref{autolag}), the lag-1 autocorrelation function
$\rho_{11}(\tau)=\langle\delta x_{1}(t+\tau)\delta x_{1}(t)\rangle$ with $\tau=1$
is calculated. Figures \ref{fig12} (a) and (b) exhibit the plots of the variance
$\sigma^{2}=\langle\delta x_{1}^{2}\rangle$, i.e. , the variance of the ComK fluctuations,
as a function of the bifurcation parameters $b_{s}$. Figures \ref{fig12} (c)
and (d) exhibit the lag-1 autocorrelation function $\rho_{11}(1)$,
associated with the ComK fluctuations, as a function of the bifurcation
parameter $b_{s}$. The parameter values used in the computations are
the same as in the case of Figure 11. In the cases of Figures \ref{fig12}(a)
and (b), the computations are carried out along the line $b_{k,1}=0.0875$
which traverses successively through the regimes of monostability, excitability,
bistability and monostability. In the cases of Figures \ref{fig12}(c) and (d),
the calculations are carried out along the line $b_{k,2}=0.15$. The shaded regions 
in Figure \ref{fig12} represent the regions of bistability with the solid (dotted) lines associated 
with low (high) ComK levels. From the plots one finds that along the line 2 ($b_{k,2}=0.15$), 
the high ComK level loses stability at the bifurcation point $b_{s,1}=0.6176$ (Figures \ref{fig12}(c) and (d)). 
The sudden regime shift from the high to the low ComK state is signaled by a diverging variance as 
the bifurcation point is approached and the lag-1 autocorrelation function attaining its maximum value 
at the point. Similar signatures are obtained at the other bifurcation point $b_{s,2}=0.7504$ where 
the low ComK state loses stability. In Figures \ref{fig12}(a) and (b), at the point $b_{s,1}=0.7799$ along 
line 1 ($b_{k,1}=0.0875$), 
as one enters a region of excitability from a region of monostability,
 the low ComK level of the monostable region continues to be stable. Since no state 
loses stability at the point, the variance $\sigma^{2}$ and the lag-1 autocorrelation 
function $\rho_{11}(1)$ do not provide any signatures of regime change. At the 
point $b_{s,2}=0.7868$, the system enters the region of bistability from a region of excitability. 
The high ComK state loses its stability at this point signaled by a diverging variance and with $\rho_{11}(1)$ 
attaining a maximum at this point. The low ComK level loses stability at the point $b_{s,3}=0.8094$ 
when the system transactions from a region of bistability to a region of monostability with high ComK 
level. Experimentally, the entry into a region of excitability/bistability from a region of monostability 
(low ComK level) is identified by the appearance of a bimodal distribution in the ComK levels as observed in 
single cell flow-cytometry measurements. In the case of excitability, however, there are 
no accompanying signatures in the measurable quantities like variance and lag-1 autocorrelation 
function. In the case of bistability, sharp rises in $\sigma^{2}$ and $\rho_{11}(1)$ indicate 
an impending regime shift.

\section{Concluding Remarks}
\label{conclude}

In this paper, we investigate the early signatures of sudden regime shifts occurring at the bifurcation points 
associated with a fold-bifurcation model. The basic concepts and methodology are applicable to a general class of 
bifurcation phenomena occurring in diverse systems. Our focus, however, is on obtaining quantitative estimates of the 
early signatures of regime shifts in the gene expression dynamics of competence development in \textit{B.subtilis}. 
While quantities like $\lambda_{max}$ and $T_{R}$ are computed with a knowledge of the deterministic rate equations, 
stochastic formalisms based on the LE, the FP equation and the LNA have been employed to calculate the other quantities.  
The early indicators of regime shifts, 
which include the CSD, variance, lag-1 autocorrelation, skewness and the ratio $r$ of MFPTs, are shown to exhibit distinctive 
variations as a bifurcation point is approached. Though the literature on the early signatures of regime shifts in ecosystems,
financial markets and complex diseases 
is extensive \cite{scheffer091,scheffer092,strange07,scheffer03,guttal08,lenton11,venegas05,mcsharry}, 
the issue has not been systematically addressed in the case of gene expression dynamics, a fundamental activity 
in the living cell. Our study is the first in this direction and illustrates how the early signatures provide knowledge in advance 
of an impending regime shift. Some of these signatures may also provide clues on the physical principles underlying gene 
expression dynamics. As already pointed out, $\lambda_{max}$ and $T_{R}$ do not exhibit any distinctive features 
(Figures \ref{fig11} (a) and (b)) when the system passes from a region of excitability (bimodal distributions in protein 
levels) to a region of monostability (unimodal distribution of low ComK levels).  
This is in
contrast to the case when the system passes from a region of bistability to one of monostability. The quantity $\lambda_{max}$, 
associated with the steady state which loses stability, becomes zero at the bifurcation point and the return time $T_{R}$ 
diverges (Figures \ref{fig11} (c) and (d)). 
Experiments detecting the CSD, as the bifurcation point is approached, are difficult to carry out. Observation of the 
CSD in recent experiments \cite{dai12,veraart12} provides pointers for carrying out further such experiments. The CSD experiment, 
if designed in the appropriate manner, would be able to distinguish between the bistability versus excitability paradigm in 
the case of competence development in \textit{B.subtilis}. Experiments based on flow-cytometry and time-lapse fluorescence 
microscopy are easier to carry out and provide estimates of the variance $\sigma^{2}$ and the lag-1 autocorrelation 
function $\rho_{11}(1)$ \cite{simpson,kaufmann}. Similar to Figure \ref{fig11}, no distinctive signatures are obtained in terms 
of $\sigma^{2}$ and $\rho_{11}(1)$ as one crosses from a region of excitability to that of monostability (Figures \ref{fig12}
(a) and (b)) whereas prominent signatures indicate the passage from bistability to monostability (Figures \ref{fig12} (c) 
and (d)). Thus flow-cytometry and time-lapse fluorescence microscopy measurements for a range of values of the bifurcation 
parameter would be able to distinguish between bistability and excitability as the physical mechanism underlying competence 
developement.

One utility of obtaining knowledge of the early signatures is that specific risk aversion strategies may be developed 
if the sudden transition is to a regime which is undesirable due to various considerations. Examples of such regime shifts 
include asthma attacks \cite{venegas05}, epileptic seizures \cite{mcsharry} and the sudden deterioration of complex diseases
\cite{chen12}. One may add the development of persistence of pathogens like \textit{M. tuberculosis} in the human lung granulomas
to the list. Recent experiments \cite{sureka08, ghosh11} on a sister species \textit{M. smegmatis} provide evidence that a 
fraction of the mycobacterial population (the population of persisters) is able to survive under nutrient depletion. To 
achieve this, the mycobacteria adopt the strategy of generating phenotypic heterogeneity in the form of two distinct 
subpopulations. The concentration of a key regulatory protein \textit{Rel} is high in one subpopulation (which develops 
into the persister subpopulation) and low in the other. In the subpopulation with high \textit{Rel} level, the stringent 
response pathway is initiated which help the mycobacteria to avoid death and adapt to nutrient depletion. The experiments 
\cite{sureka08, ghosh11} show that the principle underlying the development of phenotypic heterogeneity is based on bistability 
and noise-induced transitions between the expression states corresponding to low and high \textit{Rel} levels. The problem 
of persisters is that they are not killed by antibiotic drugs and wait for the opportune moment to restart an infection 
\cite{balaban11}. Early signatures of a regime shift from the normal to the persistent state would help in developing 
measures preventing the switch to persistence. Similar studies could be carried out on other systems in which bifurcation 
phenomena are responsible for the development of heterogeneity, choice of cell fate \cite{collins11} or regime shifts leading 
to new types of dynamical behaviour.

\appendix*

\section{A. Linear Noise Approximation}

In the Appendix A, we describe briefly the linear noise approximation
(LNA) to the Chemical Master Equation (CME) \cite{kampen92,elf} and introduce
the notations for the relevant quantities. We consider an intracellular
biochemical system with volume $\Omega$ and $N$ different chemical
components. The concentrations of the components are represented in
the form of the vector $\mathbf{x}=(x_{1},x_{2},...,x_{N})^{T}$ where
$T$ denotes the transpose. The chemical constituents take part in
$R$ elementary reactions. The state of the system is given by $\mathbf{x}$
which changes due to the occurrence of any one of the $R$ reactions.
We define the integers $S_{ij}$, $i=1,2,...,N$ , $j=1,2,...,R$
to be the elements of a stoichiometric matrix $\mathbf{S}$. The number
of molecules of the chemical components $i$ changes from $X_{i}$
to $X_{i}+S_{ij}$ when the $j$th reaction takes place.

The deterministic dynamics of the system are described by the rate
equations

\begin{equation}
\frac{dx_{i}}{dt}=\overset{R}{\underset{j=1}{\sum}\:\:}S_{ij}\,f_{j}(x)\:\:(i=1,2,...,N)
\label{arate}
\end{equation}

\noindent In a compact notation

\begin{equation}
\mathbf{\mathbf{\overset{.}{x}}=S f(x)},\:\:\mathbf{f}\left(\mathbf{x}\right)
=(f_{1}(\mathbf{x}),\cdots,f_{R}(\mathbf{x}))^{T}
\label{arate2}
\end{equation}

\noindent where \textbf{$\mathbf{f(x)}$} defines the reaction propensity
vector. In the steady state, $\mathbf{\overset{.}{x}}=0$ with the
state vector $\mathbf{x}_{s}$ determined from the condition $\mathbf{f}\left(\mathbf{x}_{s}\right)=0$.
Let $\delta\mathbf{x}$ denote a weak perturbation applied to the
steady state, i.e., the new state vector $\mathbf{x}=\mathbf{x}_{s}+\delta\mathbf{x}$.
On Taylor expansion of the rate vector $\frac{d\mathbf{x}}{dt}$ about
the steady state and retaining only terms linear in $\delta\mathbf{x}$,
one gets

\begin{equation}
\frac{d}{dt}\delta\mathbf{x}=\mathbf{A}\delta\mathbf{x}
\label{xrate}
\end{equation}

\noindent where $\mathbf{A}$ is the Jacobian matrix the elements of which
are given by

\begin{equation}
A_{ij}=\overset{R}{\underset{k=1}{\sum}}S_{ik}\:\:\frac{\partial f_{k}}{\partial x_{j}}
\label{jacob}
\end{equation}

\noindent The state $\mathbf{x}_{s}$ is stable if all the eigenvalues
of $\mathbf{A}$ have negative real parts. A deterministic dynamical model,
as described above, is appropriate for describing the dynamics of
a system when the number of molecules, $\mathbf{X}_{i}\;\;(i=1,2,...,N)$ 
is large. In reality, the biomolecules participating
in cellular reactions are mostly small in number so that a stochastic
description of the dynamics is more valid.

The CME describes the rate of change of the probability distribution
$P(X_{1},X_{2},...,X_{N},t)$ of the numbers of the different chemical
components. The CME is not exactly solvable in most cases and one
has to take recourse to various approximate methods in order to
solve the equation. The LNA provides an approximation to the CME via
a large volume ($\Omega$) expansion around the macroscopic steady
state. Noise in the form of fluctuations around the steady state is
expected to be small in the large $\Omega$ limit as the number of
the molecules scales with the volume. To the first order in the expansion,
one obtains the set of deterministic rate equations whereas in
the second order, one gets the linear FPE describing fluctuations
about the steady state. The stationary solution of the linear FPE
is given by a multivariate Gaussian distribution.
 One can further show that the covariance of the fluctuations
about the deterministic steady state is given by the fluctuation-dissipation
(FD) relation

\begin{equation}
\mathbf{A}\mathbf{C}+(\mathbf{A}\mathbf{C})^{T}+\mathbf{D}=0
\label{FD}
\end{equation}

\noindent where $\mathbf{A}$ is the Jacobian matrix (Equation (\ref{jacob})), 
$\mathbf{C}=\langle\delta\mathbf{x}\:\delta\mathbf{x}^{T}\rangle$ is
the covariance matrix, the diagonal elements of which are the variances,
and $\mathbf{D}$ is the diffusion matrix. The elements of the covariance matrix
are 

\begin{equation}
C_{ij}=\langle\delta x_{i}\:\delta xj\rangle\;\;\;\;(i,j=1,2,...,N)
\label{cov}
\end{equation}

\noindent The matrix $\mathbf{D}$ has the form

\begin{equation}
\mathbf{D}=\mathbf{S}\; \mbox{diag}(\mathbf{f}(\mathbf{x}))\;\mathbf{S}^{T}
\label{dmat}
\end{equation}

\noindent where $\mbox{diag}(\mathbf{f}(\mathbf{x}))$ is a diagonal matrix
with the elements $\mathbf{f}_{j}(\mathbf{x})$, $j=1,2,...,R$. The
matrices $\mathbf{A}$ and $\mathbf{D}$ are computed at the stationary state $\mathbf{x}=\mathbf{x}_{s}$.
Also, $\langle\delta x_{i}\rangle=0$, $i=1,2,...,N$. Once $\mathbf{A}$ and
$\mathbf{D}$ are determined, one can determine the elements of the covariance
matrix (specially, the variances) from the FD relation (\ref{FD}). The time
correlation matrix for $\delta\mathbf{x}$ is given by

\[
\langle\delta\mathbf{x}(t+\tau)(\delta\mathbf{x}(t))^{T}\rangle=\exp(\mathbf{A}\tau)\; \mathbf{C}
\]

\noindent with 

\begin{equation}
\langle\delta x_{i}(t+\tau)\delta x{}_{j}(t)\rangle
=\underset{n}{\sum}\underset{m}{\sum} e^{\lambda_{m}\tau}(T^{-1})_{mn}T_{im}C_{nj}
\label{acorrel}
\end{equation}

\noindent In Equation (\ref{acorrel}), $\lambda_{m}$'s are the eigenvalues of
the Jacobian matrix $\mathbf{A}$ and $\mathbf{T}$ is a matrix the columns of which
represent the right eigenvectors of $\mathbf{A}$.

\noindent The diagonal elements of the matrix are the autocovariances
and $\tau$ defines the lag time. The lag-$\tau$ autocorrelation
for the $i$th chemical component is

\begin{equation}
\rho(\tau)=\frac{\langle\delta x_{i}(t+\tau)\delta x{}_{j}(t)\rangle}{\sqrt{\mbox{var}\left(x_{i}(t+\tau)\right)}
\sqrt{\mbox{var}\left(x_{i}(t)\right)}}
\label{autolag}
\end{equation}

\subsection*{Acknowledgments}

MP acknowledges the support by UGC, India, vide sanction Lett. No. F.2-8/2002(SA-I) dated 23.11.2011. 
SG acknowledges the support by CSIR, India, under Grant No. 09/015(0361)/2009-EMR-I.

\end{document}